\documentclass[submission,copyright,creativecommons]{eptcs}
\providecommand{\event}{SYNT 2016}
\usepackage{breakurl}            
\usepackage{graphicx}
\usepackage{caption}
\usepackage{subcaption}
\usepackage{listings}
\usepackage[inline]{aplcomments}
\usepackage{xspace}
\usepackage{cite}
\usepackage{inconsolata}

\title{Leveraging Parallel Data Processing Frameworks with Verified Lifting}
\author{
Maaz Bin Safeer Ahmad
\institute{Computer Science \& Engineering\\
University of Washington
}
\email{maazsaf@cs.washington.edu}
\and
Alvin Cheung
\institute{Computer Science \& Engineering\\
University of Washington
}
\email{akcheung@cs.washington.edu}
}
\def\titlerunning{Leveraging Parallel Data Processing Frameworks with Verified Lifting}
\def\authorrunning{M. B. S. Ahmad \& A. Cheung}

\newcommenter{ac}{0.4,1.0,1.0}
\newcommenter{ma}{1.0,0.4,1.0}

\newcommand\sys{\textsc{Casper}\xspace}

\newcommand{\figlabel}[1]{\label{f:#1}}
\newcommand{\seclabel}[1]{\label{s:#1}}
\newcommand{\tablabel}[1]{\label{t:#1}}
\newcommand{\eqlabel}[1]{\label{eq:#1}}
\newcommand{\secref}[1]{\S\ref{s:#1}}  % for use in text
\newcommand{\figref}[1]{Figure~\ref{f:#1}}     % for use in text
\newcommand{\tabref}[1]{Table~\ref{t:#1}}  % for use in text
\newcommand{\eqref}[1]{Equation~\ref{eq:#1}}  % for use in text

\renewcommand{\sp}{\  \; | \; \ }
\newcommand{\andsp}{\ \; \wedge \; \ }
\newcommand{\forallsp}{\ \forall \;\; \ }
\newcommand{\tabsp}{\ \hspace*{0.5cm} \ }

\definecolor{javapurple}{rgb}{0.5,0,0.35}
\definecolor{javagreen}{rgb}{0,0.4,0}
\lstdefinelanguage{java}
{
  morekeywords={for, int, static, class, public, void, if, new, return},
  morecomment=[l]{//}, % l is for line comment
  morecomment=[s]{/*}{*/}, % s is for start and end delimiter
  morestring=[b]", % defines that strings are enclosed in double quotes
  basicstyle=\footnotesize\ttfamily,
  escapeinside={@}{@},
  numbers=left, firstnumber=1, numberstyle=\tiny\color{gray},
  showstringspaces=false,
  keywordstyle=\color{javapurple},
  columns=fullflexible
}

\begin{document}
\maketitle
\begin{center}
\vspace{-0.275in}
{\tt \footnotesize{http://casper.uwplse.org}}
\vspace{0.1in}
\end{center}

\begin{abstract}
Many parallel data frameworks have been proposed in recent years 
that let sequential programs access parallel processing. 
To capitalize on the benefits of such frameworks, existing code must often 
be rewritten to the domain-specific languages that each
framework supports. This rewriting---tedious and error-prone---also requires 
developers to choose the framework that best optimizes performance given a 
specific workload.

This paper describes \sys, a novel compiler that automatically retargets
sequential Java code for execution on Hadoop, a parallel data processing framework that 
implements the MapReduce paradigm. Given a sequential code fragment, \sys uses 
{\em verified lifting} to infer a high-level summary expressed in our program 
specification language that is then compiled for execution on Hadoop. We demonstrate 
that \sys automatically translates Java benchmarks into Hadoop. The translated 
results execute on average $3.3\times$ faster than the sequential 
implementations and scale better, as well, to larger datasets.
\end{abstract}

\section{Introduction}
As computing becomes increasingly ubiquitous, storage cheaper, and data 
collection tools more sophisticated, more data is being collected today than
ever before. Data-driven advances are increasingly prevalent in various scientific
domains. As such, effectively analyzing and processing huge datasets poses a grand 
computational challenge.

Many parallel data processing frameworks have been developed to handle very large 
datasets~\cite{hadoop,spark,storm,mongodb,graphlab}, and new ones continue to be 
frequently released~\cite{mongodb,tensorflow,dataflow}.
Most parallel data processing frameworks come with domain-specific optimizations 
that are exposed either via library 
APIs~\cite{hadoop,spark,storm,graphlab,tensorflow,dataflow} or high-level, 
domain-specific languages (DSLs) for users to express their computations~\cite{mongodb,halide}.
Computations expressible using such API calls or DSLs are more efficient
thanks to the frameworks' domain-specific optimizations~\cite{hive,delite,halide,stencils}.

However, the many issues with this approach often make domain-specific
frameworks inaccessible to non-experts such as researchers studying physical or
social sciences. 
First, domain-specific optimizations for different workloads 
require an expert to decide up front the most appropriate framework for a given
piece of code. Second, end users must often learn new APIs or 
DSLs~\cite{hadoop,spark,storm,graphlab,tensorflow,dataflow} and rewrite existing 
code to leverage the benefits provided by 
these frameworks. Doing so requires not only significant time and 
resource but also risks introducing new bugs into the application.
Moreover, even users willing to rewrite their applications must first
understand the intent of the code which might have been written by others.
And manually written, low-level optimizations in the code often obscure 
high-level intent. Finally, even after learning new APIs and rewriting code, 
newly emerging frameworks often turn freshly rewritten code into legacy 
applications. Users must then repeat this process to keep pace with new 
advances, requiring significant time investments that could be better spent in 
advancing scientific discovery.

One way to improve the accessibility of these parallel data processing frameworks
involves building compilers that automatically convert applications written in 
common general-purpose languages (such as 
Java or Python) to high-performance parallel processing frameworks, such as 
Hadoop or Spark. Such compilers let users write their applications in familiar 
general-purpose languages and let the compiler retarget portions of 
their code to high-performance DSLs~\cite{mold,qbs,stng}. 
The applications can then leverage the performance of these specialized frameworks without 
the overhead of learning how to program individual DSLs. 
But such compilers don't always exist, and building one can prove highly 
complex.

This paper demonstrates the application of {\em verified lifting} to automatically
convert sequential Java code fragments to MapReduce. As input, verified lifting 
takes program fragments written in a general-purpose language and uses program 
synthesis to {\em automatically find} 
provably correct code summaries. These summaries---expressed 
in our program specification language---encode the semantics of the input code 
fragment. The found summaries are then used to translate the original input code to the 
target high-performance DSL. 

The concept of verified lifting has been previously 
applied to database applications~\cite{qbs} and stencil computations 
\cite{stng}. This paper applies verified lifting to the conversion of 
sequential data processing Java code to leverage the parallel data processing 
frameworks Apache Hadoop. The problem statement remains familiar
and was first proposed in the MOLD compiler~\cite{mold}, which translates sequential 
Java code for execution on Apache Spark. MOLD uses pre-defined rewrite rules to search 
the space of equivalent Apache Spark implementations.
It scans the input code for patterns that trigger such 
rewrite rules, an approach fraught
with many limitations. For instance, it requires the a priori definition of complicated 
rewrite rules, which can be extremely brittle to code pattern changes. 
In comparison, our approach analyzes program {\em semantics} rather 
than program {\em syntax}, making it robust to code pattern changes. We also do 
not rely on pre-defined translation rules and can thus discover new solutions 
and optimizations that the user never knew existed.

We implemented our approach described above in a compiler called \sys. 
By converting sequential code fragments to Hadoop, \sys parallelizes computation 
at crucial program points where input data collections are being processed. We 
used \sys to convert five benchmark programs with encouraging results. This paper 
thus makes the following contributions:

\begin{itemize}  
	\item We describe the use of {\em verified lifting} to retarget sequential 
	Java applications to Hadoop by converting code fragments within the 
	application to Hadoop MapReduce tasks.
	
  	\item We design a {\em new program specification language} to express the intent of 
  	Java code fragments using the MapReduce paradigm.
	
	\item We employ {\em static program analysis techniques} to 
	intelligently restrict the search space of all possible summaries that can 
	be expressed in our specification language and use {\em inductive synthesis} 
	to find provably correct summaries for each input code fragment.
	
	\item We present encouraging preliminary results from using 
	\sys to identify and optimize code fragments written in sequential Java.
	To show the potential of our approach, we evaluate our system on five 
	MapReduce benchmarks used in prior work~\cite{pheonix} to demonstrate its capabilities and 
	limitations.
\end{itemize}

In the following, we describe \sys's design and 
illustrate its use to convert sequential Java programs into Hadoop tasks in~\secref{systemOverview}. 
In~\secref{description} we explain verified lifting and describe how we 
implemented each of its steps in \sys. 
~\secref{evaluation} evaluates how \sys performs using varied
benchmarks and shares our preliminary results. \secref{relatedWork} describes
related work, and we conclude in~\secref{conclusion}.

\begin{figure}[t]
    \centering
    \includegraphics[height=11cm]{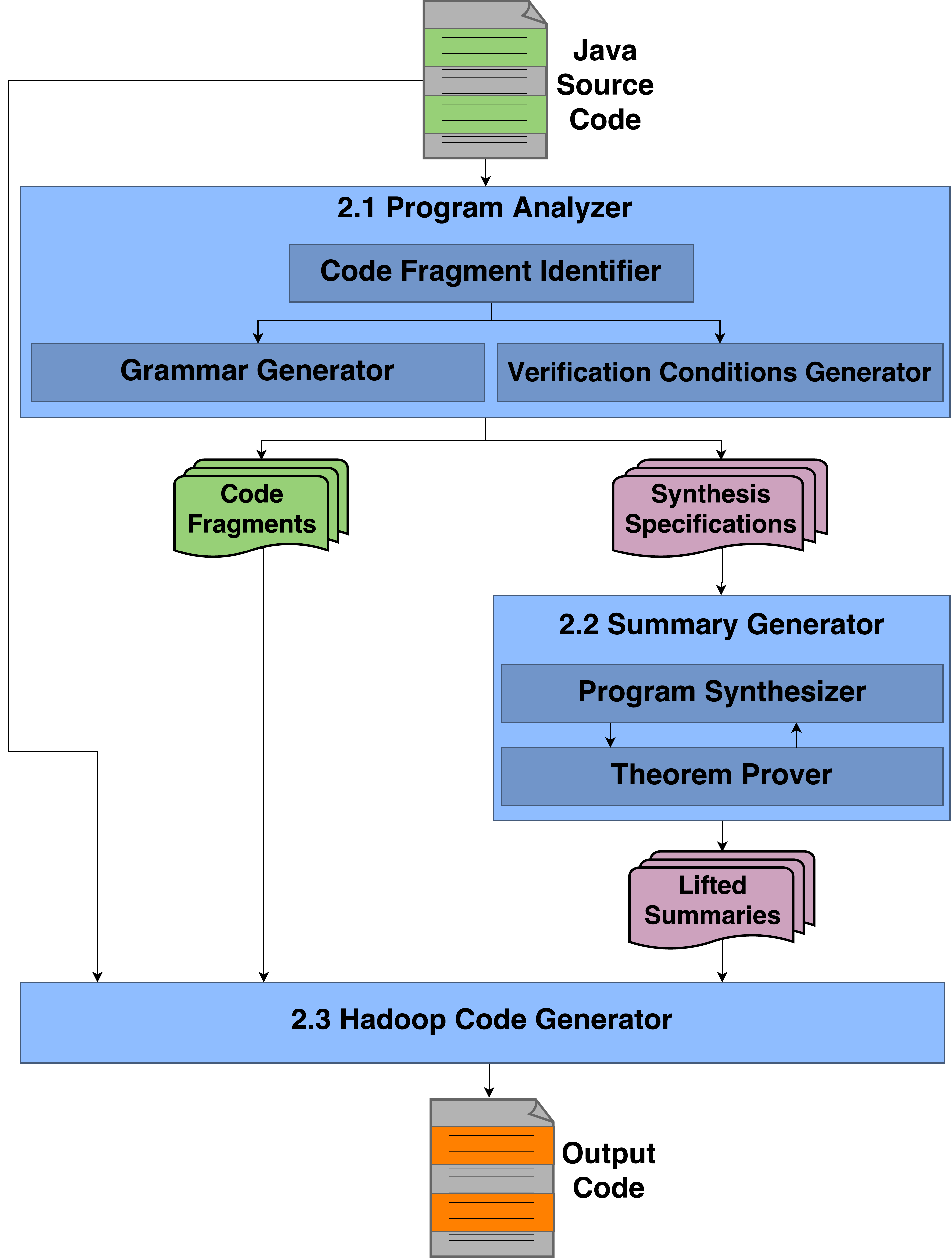}
    \caption{\sys system architecture diagram. Sequential code fragments (highlighted 
    green) in the input source file are translated to equivalent Hadoop tasks (highlighted
    orange).}
    \figlabel{arch}
\end{figure}

\begin{figure}[t]
\centering
\begin{subfigure}[b]{0.49\textwidth}
\begin{subfigure}[t]{\textwidth}
\begin{small}
\begin{lstlisting}[language=java]
int[][] histogram(int[] data) {
  int[] hR = new int[256];
  int[] hG = new int[256];
  int[] hB = new int[256];
  for (int i = 0; i < data.length; i += 3){
    int r = data[i];
    int g = data[i + 1];
    int b = data[i + 2];
    hR[r]++; @\label{lst:countStart}@
    hG[g]++; 
    hB[b]++; @\label{lst:countEnd}@
  }
  int[][] result = new int[3][];
  result[0] = hR;
  result[1] = hG;
  result[2] = hB;
  return result;
}
\end{lstlisting}
\end{small}
\vspace{-0.1in}
\caption{Input source code}
\figlabel{full_example_input}
\end{subfigure}
\begin{subfigure}[b]{\textwidth}
\begin{small}
\begin{lstlisting}[language=java,numbers=none]
Map kvPairs = HistogramHadoop.execute();
hR = kvPairs.get(0); 
hG = kvPairs.get(1);
hB = kvPairs.get(2);
\end{lstlisting}
\end{small}
\caption{\sys wrapper to replace the loop in (a)}
\figlabel{full_example_output}
\end{subfigure}
\end{subfigure}
~
\begin{subfigure}[b]{0.49\textwidth}
\begin{small}
\begin{lstlisting}[language=java]
public class HistogramHadoop{
  class HistogramMapper extends Mapper {
    void map(int key, int[] value){
      for (int i=0; i<value.length; i+=1) {
        if(i%3==0) emit((0, value[i]), 1); @\label{lst:emitStart}@
        if(i%3==1) emit((1, value[i]), 1);
        if(i%3==2) emit((2, value[i]), 1); @\label{lst:emitEnd}@
  }}}
  class HistogramReducer extends Reducer {
    void reduce(Tuple key, int[] values) {
      int value = 0;
      for (int val:values) {value=value+val;} @\label{lst:reduce}@
      emit(key, value);
  }}
  static Map execute() {
    Job = Job.getInstance();
    job.setMapper(HistogramMapper);
    job.setReducer(HistogramReducer);
    return job.execute();
  }
} 
\end{lstlisting}
\end{small}
\caption{\sys-generated Hadoop task}
\figlabel{full_example_mr}
\end{subfigure}
\vspace{-0.1in}
\caption{\sys translation of the 3D Histogram benchmark.}
\vspace{-0.1in}
\figlabel{full_example_hist}
\end{figure}

\section{System Overview}
\seclabel{systemOverview}

This section describes the architecture of the \sys compiler. 
\sys automatically identifies and 
converts sequential Java source code fragments into semantically equivalent 
MapReduce tasks implemented using Hadoop. It generates new, optimized version of the 
input source code where original code fragments are replaced by invocations to 
the generated MapReduce tasks. \figref{arch} shows \sys's different components 
and how they interact in the compilation pipeline.

Before explaining each component in detail, we should generally note that by
statically analyzing Java input source code, \sys extracts code fragments that can
potentially be translated. It then generates a high-level summary of each 
extracted code fragment. Expressed in our high level program specification
language (see \secref{verifiedlifting}), the summary is inferred by a program 
synthesizer. To quickly traverse the large space of possible summaries,
\sys bounds the search space considered by the synthesizer and uses a bounded
model-checking procedure to locate any candidate candidate summaries for a given
code fragment.\footnote{While not yet implemented in the current 
\sys prototype, any candidate summary that passes the bounded model checking
phase will be forwarded to a theorem prover, which verifies that the 
synthesizer-generated summary is semantically equivalent to the original code.}  
The Hadoop code generator module uses the verified summary to 
produce code for Hadoop MapReduce tasks. Lastly, the code generator module
prepares a new version of the input source code by replacing the original code
fragments with invocations to the translated Hadoop tasks.

Throughout the remainder of this paper, we use, as a running example, a benchmark
from the Phoenix suite of benchmarks~\cite{pheonix} that generates 3D histograms 
from image data stored in a file. \figref{full_example_hist} shows \sys's translation 
of this 3D Histogram benchmark. In~\figref{full_example_input}, 
the original program 
sequentially iterates over an array of integers 
representing the intensity values of colors red, green, and blue for each pixel. 
It then counts the number of times each value occurs for each color from Lines~\ref{lst:countStart} to
\ref{lst:countEnd}. The parallel program \sys generates, on the other hand, emits 
a key-value pair with the tuple (color, intensity) as key and 1 as value (lines~\ref{lst:emitStart} to
\ref{lst:emitEnd} in~\figref{full_example_mr}). The generated pairs are then grouped
by key and the frequency of each key is calculated by adding
all the 1's in the reducer phase (line~\ref{lst:reduce}). \figref{full_example_output}
shows the final code generated by \sys that replaces the original loop, where it
invokes the Hadoop
task shown in~\figref{full_example_mr} and uses the output from the Hadoop task
to update program state accordingly. In our evaluation, the translated version of
the benchmark performed over $1.5\times$ faster for a dataset of 50GB.

We now discuss the three essential modules that make up \sys's compilation pipeline.

\subsection{Program Analyzer}
\seclabel{programAnalyzer}
The program analyzer, the first component in \sys's compilation pipeline, has two
goals. First, it identifies all code fragments that are candidates for conversion.
Second, it generates synthesis specifications for each identified 
candidate code fragment. The program analyzer operations are grouped into 
three sub-components: the code fragment identifier, the verification conditions
generator, and the grammar generator.

In the current prototype, \sys's code fragment identifier finds loops from the 
input program and 
extracts them as candidates for conversion. \sys currently does not consider 
non-looping code fragments (such as recursive functions) as candidates for conversion. 
In addition, \sys also 
ignores loops containing calls to external library methods that 
are unrecognized by the \sys compiler. In \secref{codeExtraction}, we formally list
the criteria needed for a code fragment to be extracted as a candidate for conversion, highlighting
some current limitations of \sys's implementation.

The second program analyzer sub-component is the grammar generator, which aims
to confine the space of synthesizable summaries. Without doing so, the space of all 
possible summaries expressible in our program specification language would be too large.
The grammar generator takes as input the code fragments extracted by the code 
fragment identifier and statically analyzes each one to extract semantic information. 
It then uses the extracted information to generate a program grammar for every 
code fragment. The challenge here is to generate a grammar expressive enough to
express the correct summary, but not so expressive as to make the problem of
summary search intractable.
In \secref{grammarGeneration}, we explain how the 
grammar generator leverages static program analysis to construct a grammar for 
each code fragment.

The third program analyzer sub-component is the verification conditions 
generator. This component uses Hoare logic~\cite{hoare} and static
program
analysis to generate verification conditions for each code fragment.
{\em Verification conditions} are logical statements that describe what must be 
true for a given summary to be semantically equivalent to the 
original code fragment. We explain how \sys uses Hoare style program verification to verify program 
equivalence in \secref{verifyingEquivelance}.

The output of the verification conditions generator is a search template for the 
summary, with: (i) the search space specified by the grammar
generator, and (ii) the verification conditions generator producing the logical 
assertions that must be satisfied given a candidate summary. The summary generator
uses this template to search for a valid summary of the input code 
fragment, as we next explore.

\subsection{Summary Generator}
Using the program analyzer's specifications, the summary generator traverses 
the search space to find a summary that satisfies the verification conditions. 
It consists of two modules: the program synthesizer and the theorem prover. The 
synthesizer takes the search space description and verification conditions 
previously generated and searches for a code summary that satisfies the verification 
conditions. To make the search problem tractable, it uses a bounded model checking 
procedure: the synthesizer checks for correctness only over a small sub-domain. 
When a promising candidate for the summary is found, viz., one that satisfies 
the verification conditions in the sub-domain, it is passed onto 
the formal theorem prover~\footnote{See footnote 1.}, 
which checks it for correctness over the entire 
domain of inputs. Candidate solutions that fail the formal verification step are eliminated from the 
search space, and the search restarts for a new candidate solution. Using this 
two-step verification process helps \sys quickly discard bad candidates. 
The more computationally expensive process of formal verification is reserved 
only for promising candidate solutions. As output, the summary generator emits a 
verifiably correct summary for each code fragment that can then be translated to Hadoop. 
Note that the summary generator may not always find a solution that can be proven correct:
some fragments are impossible to translate to MapReduce while others 
might have complex solutions that \sys currently can not generate. In such cases, \sys 
gives up on translating the code fragment.

\subsection{Hadoop Code Generator}
The summaries synthesized by the summary generator are expressed in our high-level
program specification language. Generating Hadoop implementations from these high level 
program specifications is straightforward and is achieved in \sys through syntax-directed
translation rules. The code generator module also outputs the code required
to embed these Hadoop tasks into the original program. Essentially, \sys generates 
a new (source) version of the input code, where each code fragment that was 
successfully translated is replaced by code that first invokes the 
corresponding Hadoop task and then uses the output generated by the Hadoop task
to update the state of the program. \figref{full_example_output} shows such 
generated wrapper code for the 3D Histogram example. We present more details 
about \sys's code generation module in \secref{codegen}.

\section{Converting Code Fragments}
\seclabel{description}
This section explains how \sys uses verified lifting to convert sequential
Java code fragments to MapReduce tasks. We review the concept of verified 
lifting in \secref{verifiedlifting} and describe the program specification 
language \sys uses to express program summaries. In \secref{verifyingEquivelance}, we
explain how \sys verifies that the identified summaries preserve program semantics of
the original code fragment. \secref{specgeneration} discusses the search
process \sys uses to find program summaries, while \secref{codeExtraction} explains
how \sys selects suitable code fragments for translation. Finally, \secref{codegen}
explains code generation after the program summary has been inferred.

%%%%%%%%%%%%%%%%%%%%%%%%%%%%%%%%%%%%%%%%%%%%%%%%%%%%%%%%%%%%%%%%%%%%%%%%%%%%%%%%

\subsection{Verified Lifting}
\seclabel{verifiedlifting}
Verified lifting~\cite{stng,qbs} is a general technique that infers the semantics of 
code written in a general-purpose language by ``lifting'' it to summaries expressed
using a high-level language. \sys specifies code fragment summaries in our program
specification language in the form of postconditions that describe the effects of the
code fragment on its {\it output variables}, i.e., variables that are modified within 
the code fragment. The goals of our program specification language are:

\begin{itemize}
	\item To generate summaries that \sys can translate to the target
  platform DSL. This excludes valid summaries that cannot be translated. Therefore,
  the language should omit constructs that cannot be translated
  easily to the target.

	\item To generate non-trivial summaries that exhibit parallel 
  data processing. Obviously, this excludes summaries that execute the 
  computation sequentially. \secref{parallelism} discusses
	the sources of parallelism in MapReduce and how \sys generates 
	solutions that exploits them.
\end{itemize} 

With these goals in mind, \sys's inferred summaries must be of the 
form:
\begin{eqnarray}
\forall~v~\in~outputVariables~.~v~=~reduce(map(data,f_m),f_r)[id_v]
\eqlabel{postcond}
\end{eqnarray}
where $data$ is the iterable input data collection. The $map$ function 
iterates over the $data$ while calling the $f_m$ function on each element.
$f_m$ takes as input an element from $data$ and generates potentially multiple 
key-value pairs. $map$ then collects and returns key-value pairs generated by 
invocations of $f_m$. The $reduce$ function takes these key-value pairs, groups 
them by key, and calls $f_r$ for each key and all values that 
correspond to that key. Function $f_r$ aggregates all values for the
given key and emits a single key-value pair. Like $map$, $reduce$ collects all 
aggregated key-value pairs and returns an associative array that maps each 
variable's ID to its final value. The variable ID is a unique identifier that
\sys assigns to every output variable. \sys requires that summaries (i.e., postconditions) 
be of the form described in Eqn.~(1) for easy translation to Hadoop tasks.

% this is discussed in 3.4.1
%In practice, \sys focuses on loops since they can most likely be converted to
%Hadoop; input variables correspond to variables declared outside the loop and 
%read inside the loop body. Similarly, output variables are modified inside the 
%loop body but not declared there.

In the preceding discussion, $f_m$ and $f_r$ remain unspecified. Verified lifting 
seeks a definition of $f_m$ and $f_r$ that makes a valid inferred summary, 
viz., one that preserves the semantics of the input code fragment. To do this in
\sys, the synthesizer generates the implementation 
of these two functions (see \secref{specgeneration}) using the verification 
conditions computed by the program analyzer for each code fragment 
(see \secref{codeExtraction}).

%%%%%%%%%%%%%%%%%%%%%%%%%%%%%%%%%%%%%%%%%%%%%%%%%%%%%%%%%%%%%%%%%%%%%%%%%%%%%%%%

\subsection{Verifying Equivalence}
\seclabel{verifyingEquivelance}
The summaries \sys infers must be semantically equivalent to the 
input code fragment. \sys establishes the validity of the inferred 
postconditions using Hoare-style verification conditions~\cite{hoare}, which 
represent the weakest preconditions of a code fragment that must be true to 
establish the postcondition of the same code fragment under 
all possible executions. Generating verification conditions for simple 
assignment statements and conditionals is straightforward. For example, consider the 
imperative program statement {\tt x := y + 3}. To show that the candidate 
postcondition {\tt x > 10} is a valid postcondition, we must prove that 
{\tt y + 3 > 10} is true before the statement is executes. In this case, 
{\tt y + 3 > 10} is called the {\em verification condition} for this postcondition. 
Computing verification 
condition is easy for simple statements.
For a loop, however, computing verification conditions becomes more difficult 
since a loop invariant is needed. The {\em loop invariant} is a hypothesis that 
asserts that the  postcondition is true regardless of how 
many times the loop iterates. Hoare logic states that the following three 
statements must hold for the loop invariant (and postcondition) to be valid:

\begin{enumerate}  
	\item $\forall \sigma. \; preCondition(\sigma) \rightarrow loopInvariant(\sigma)$
	\item $\forall \sigma. \; loopInvariant(\sigma) \; \wedge \; loopCondition(\sigma)  \rightarrow loopInvariant(body(\sigma))$
	\item $\forall \sigma. \; loopInvariant(\sigma) \; \wedge \; \neg loopCondition(\sigma)  \rightarrow postCondition(\sigma)$
\end{enumerate}

Statement 1 asserts that the loop invariant must be true when the precondition 
is true for all program states ($\sigma$), i.e., the loop invariant
must be true before entering the loop. Statement 2 asserts that for all 
possible program states $\sigma$---assuming that the loop invariant is true and that the 
loop continues---the loop invariant remains true after one more execution of
the loop body; (here, $body(\sigma)$ returns a new program state after executing 
the loop body at $\sigma$). Statement 3 asserts that if the loop invariant is 
true and if loop terminates, then the postcondition must be true for all
possible program states.

Two challenges affect the identification of postconditions (and hence
summaries) for code fragments that involve loops. 
First, {\em both} the loop invariants and postcondition must be synthesized. 
Unlike prior work on searching for invariants~\cite{daikon,templateinv},
however, \sys needs to find loop invariants that are only logically strong enough 
to establish the soundness of the postcondition, i.e., 
those that satisfy statement 3. This is made easier thanks
to the specific form of the postcondition that \sys looks for.
In addition, establishing the validity of 
the found invariants and postconditions requires checking {\it all} possible 
program states, complicating the synthesis problem. We discuss
how \sys makes the search problem manageable in \secref{sproced}.

%%%%%%%%%%%%%%%%%%%%%%%%%%%%%%%%%%%%%%%%%%%%%%%%%%%%%%%%%%%%%%%%%%%%%%%%%%%%%%%%

\subsection{Searching for summaries}
\seclabel{specgeneration}
\sys seeks to infer a summary for each code fragment, where each 
summary is a postcondition of the form explained in \secref{verifiedlifting}. 
This section describes how \sys uses synthesis to search for postconditions 
{\it and} the loop invariants they require to prove the postconditions correct.

\subsubsection{Generating Verification Conditions}
\seclabel{templateGeneration}
In \secref{verifyingEquivelance}, we explained the three verification conditions
that must be satisfied by the synthesized summary. These verification conditions
involve a precondition, postcondition, and loop invariant for the code fragment. 
Preconditions are generated by extracting, through static program analysis, 
the program state (values of input and output variables) just before the 
loop starts executing. When the value of a variable before the loop starts cannot 
be determined, \sys generates a new variable to represent the initial value. 
The loop invariant has a form similar to the postcondition (see \secref{verifiedlifting});
unlike the postcondition, however, which calls $map$ and $reduce$ on the entire data 
collection, the loop invariant calls $map$ and $reduce$ only on the subset of the 
collection that has so far been traversed by the loop. Also, the loop invariant 
includes an expression that describes the behavior of the loop counters.

\begin{figure}
\centering
\begin{subfigure}[t]{0.80\textwidth}
\begin{small}
\(
preCondition(hR, hG, hB, i) \equiv \\
\tabsp hR = [0..0] \andsp hG = [0..0] \andsp hB = [0..0] \andsp i = 0 \\[0.2cm]
postCondition(data, hR, hG, hB) \equiv \\
\tabsp \forallsp 0 \leq j < hR.length. \;\; hR[j] = reduce(map(data, f_m), f_r)[(0,j)] \;\; \wedge \\
\tabsp \forallsp 0 \leq j < hG.length. \; hG[j] = reduce(map(data, f_m), f_r)[(1,j)] \;\; \wedge \\
\tabsp \forallsp 0 \leq j < hB.length. \;\; hB[j] = reduce(map(data, f_m), f_r)[(2,j)]  \\[0.2cm]
loopInvariant(data, hR, hG, hB, i) \equiv \\
\tabsp LoopCounterExp \; \wedge \\
\tabsp \forallsp 0 \leq j < hR.length. \;\; hR[j] = reduce(map(data[0:i], f_m), f_r)[(0,j)] \;\;  \wedge  \\
\tabsp \forallsp 0 \leq j < hG.length. \; hG[j] = reduce(map(data[0:i], f_m), f_r)[(1,j)] \;\; \wedge  \\
\tabsp \forallsp 0 \leq j < hB.length. \;\; hB[j] = reduce(map(data[0:i], f_m), f_r)[(2,j)] 
\)
\vspace{0.1cm}
\end{small}
\end{subfigure}
\caption{Definitions of precondition, postcondition and loop invariant for the 3D Histogram example.}
\figlabel{hist_summary_template}
\end{figure}

\figref{hist_summary_template} shows the precondition, postcondition and loop 
invariant generated for the 3D Histogram benchmark. The postcondition and loop 
invariant functions describe the behavior that must be true for the bodies of 
$f_m$ and $f_r$ to be correct. For example, the postcondition states that for 
each index $j$ of $hR$, the value of $hR[j]$ must equal to the output of map and 
reduce functions for key $(0,j)$.

\subsubsection{Specifying Search Space}
\seclabel{grammarGeneration}
This section describes how \sys generates the grammar that the synthesizer uses to 
construct bodies of $f_m$ and $f_r$. By dynamically generating a grammar for 
each code fragment, \sys restricts the space of summaries through which the 
synthesizer must search.

Recall that the function $f_m$ takes as parameters the input data collection 
and an index into the collection and returns a set of key-value pairs. \sys 
constructs the body of $f_m$ using emit statements and conditionals. The current
\sys prototype does not generate implementations of $f_m$ that involve loops. Based on
our experiments, we found that 
using the same number of emit statements as output variables 
in the code fragment works well as a starting point. 
The number of emit statements can then be increased if a solution cannot be 
found. In general, however, \sys takes a conservative approach to avoid 
implementations with redundant emit statements since they generate unnecessary 
shuffle data, consequently hurting performance. Each emit 
statement produces a key-value pair; the key and value can be any expression
generated by one of our expression grammars or tuples of such expressions.

The $f_r$ function reduces all values emitted by $map$ for a given key 
into a single value. The body of $f_r$ implements the folding operation. \sys 
uses the synthesizer to generate the folding expression that reduces two values 
into one. It also generates an expression grammar to synthesize the folding expression.

\sys generates expression grammars for each primitive data type found in the code
fragment. Each grammar
can be used to generate expressions that evaluate to a value of its type.
The expressions are formulated using the operators and function calls from the 
original code fragment. Input variables, loop counters, and literals from the code 
fragment are used as terminals. For arithmetic types, \sys lets the synthesizer  
generate new constants as well. Furthermore, \sys generates an expression grammar
to construct the folding expression in $f_r$ and the loop counter expression in 
the loop invariant.

All expression grammars generated by \sys are bounded to a set level of recursion which the 
user can specify. The recursive bound of a grammar controls the amount of times
the synthesizer is allowed to expand the non-terminals while formulating an expression.
If the synthesizer cannot find a solution, the expression grammars can be
incrementally expanded by either introducing new operators and functions 
that were not found in the code fragment or increasing the recursive bound on the
grammar. The order in which new constructs are added to the grammar is guided by 
priority values that we have encoded into \sys.

\figref{hist_summary_grammar} shows the grammar generated for the 3D Histogram
benchmark after 2 iterations of grammar expansion. It is easy to see how to the solution 
presented in \figref{full_example_hist} can be generated from this grammar.

\begin{figure}
\centering
\begin{subfigure}[t]{0.85\textwidth}
\begin{small}
\centering
\begin{eqnarray*}
\mathsf{ f_m} &::=&  \mathsf{\{ EmitMap; \;\; EmitMap; \;\; EmitMap; \}} \\
\mathsf{EmitMap} &::=& \mathsf{emit(Exp,\;Exp) \sp if(BoolExp)\{ \; emit(Exp,\;Exp) \; \}} \\
\mathsf{Exp} 	 &::=& \mathsf{IntExp \sp BoolExp \sp (Exp, Exp)} \\
\mathsf{IntExp}  &::=& \mathsf{IntTerm \sp data[IntExp] \sp IntExp \; + \; IntExp \sp IntExp \; \% \; IntExp }\\
\mathsf{IntTerm} &::=& \mathsf{intLiteral \sp loopCounter }\\
\mathsf{BoolExp} &::=& \mathsf{true \sp false \sp IntExp == IntExp \sp BoolExp \;\; \wedge \;\; BoolExp }\\
				 &\sp& \mathsf{BoolExp \;\; \vee \;\; BoolExp }\\
\mathsf{f_r} &::=&  \mathsf{\{ value = IntLiteral; \;\; for(v \;\; in \;\; values) \{ \; value = FoldExp \; \} \;\; emit(key,value); \} }\\
\mathsf{FoldExp} &::=& \mathsf{FoldTerm \sp FoldExp \; + \; FoldExp }\\
\mathsf{FoldTerm} &::=& \mathsf{intLiteral \sp value \sp v }\\
\mathsf{LoopCounterExp} &::=& \mathsf{LoopTerm \;\; <= \;\; LoopTerm \;\; <= \;\; LoopTerm }\\
\mathsf{LoopTerm} &::=& \mathsf{loopCounter \sp intLiteral \sp data.length}
\end{eqnarray*}
\end{small}
\end{subfigure}
\vspace{-0.5cm}
\caption{Grammar generated for 3D Histogram example.}
\figlabel{hist_summary_grammar}
\end{figure}

\subsubsection{Search Procedure}
\seclabel{sproced}
Despite all the search space constraints already discussed, the space of 
possible summaries remains large. Therefore, to accelerate the search, 
\sys splits the verification process into two parts: it first uses a bounded-checking 
procedure to find candidate invariants and postconditions. For candidate invariants 
and postconditions that pass the bounded-checking procedure, it then uses a 
theorem prover to establish soundness for all input program states. If the 
theorem prover fails (via a timeout) or returns unsat, the synthesizer continues to search 
for a new candidate summary in the same search space. When it finds no more 
candidate summaries, the synthesizer expands the grammar to increase the search 
space. It does this by either adding new non-terminals, increasing the recursive 
bound for the grammar, or increasing the number of emits made by $f_m$, as discussed 
earlier. Configuration parameters specified by the user control this iterative
expansion of the search space. Eventually, the synthesizer either finds a 
verifiably correct summary or halts efforts to convert the code fragment. 

\sys also decouples the synthesis procedure from formal verification and uses 
off-the-shelf tools for each of the two sub-problems. This methodology works well 
in practice to reduce the synthesis time.

%%%%%%%%%%%%%%%%%%%%%%%%%%%%%%%%%%%%%%%%%%%%%%%%%%%%%%%%%%%%%%%%%%%%%%%%%%%%%%%%

\subsection{Initial Code Extraction}
\seclabel{codeExtraction}
The current \sys prototype parses the abstract syntax tree (AST) of the input 
program source code to extract loops as individual fragments. \sys then analyzes 
each fragment's AST to ensure it meets the following criteria:

\begin{itemize}
	\item The code fragment contains no unsupported library function calls. To 
	synthesize summaries, \sys must identify input and output variables 
	(see \secref{extractvars}), and the lack of library source code makes this 
	impossible unless models that describe library function semantics are encoded into 
	the compiler. \sys currently supports commonly used library functions, such 
	as methods of the {\tt java.lang.\{String,Integer\}} and 
	{\tt java.util.\{ArrayList,Map\}} classes.

	\item Each loop contains no unstructured control flow. \sys's current 
	implementation cannot extract necessary semantics from such loops, such as 
	the premature terminations and loop stride.

  	\item The code fragment contains no nested loops. \sys does not currently 
  	process nested loops. If any is found, \sys attempts to	optimize only the 
  	innermost loop.

	\item The code fragment contains no assignment statements that can create an 
	alias. Moreover, \sys does not currently perform any alias analysis and 
	assumes that none of the input variables in the code fragment is aliased. 
	Thus, user defined objects cannot be assigned. Fields of these objects 
	can be modified as long as they are a primitive type. Similarly, array 
	indexes can be modified---if array is of an immutable type---but not the 
	pointer to an array. Support for assigning common immutable data structures, 
	such as {\tt java.lang.\{Integer,String\}}, has been built into the compiler. 

\end{itemize}

\sys overlooks code fragments that do not satisfy these criteria. Once a loop has 
been marked for conversion, it is normalized to a simpler form before further 
analysis. The normalization breaks down large instructions 
into smaller, simpler ones (such as breaking down all expressions into binary ones)
and converts all loop constructs into 
{\tt while(true)\{...\}} loops. All of which are standard compiler transformations.

\subsubsection{Extracting Input and Output Variables}
\seclabel{extractvars}
\sys makes additional passes on the normalized AST to extract input and output 
variables. It examines each assignment statement inside the code fragment in 
isolation and extracts assignment targets as output variables. Similarly, 
all variables in the source of an assignment are extracted as input 
variables (this may also include some output variables). Local variables declared 
inside a loop body are considered neither 
input nor output variables. To determine whether a function call parameter is an 
input or output variable, \sys must analyze the function's source code. For 
library functions, this information must be encoded into \sys beforehand. If a 
constant index of an array is accessed, then a separate input variable is 
created for the array element. However, if any dynamic accesses are made, then the 
entire array is considered an input variable.

For the 3D Histogram example shown in~\figref{full_example_hist}, arrays {\tt hR}, {\tt hG} and 
{\tt hB} are labeled as output variables, and the {\tt data} array is identified as an 
input variable. Variables {\tt i}, {\tt r}, {\tt g} and {\tt b}---all declared inside the 
loop body---are considered to be neither input nor output variables.

%%%%%%%%%%%%%%%%%%%%%%%%%%%%%%%%%%%%%%%%%%%%%%%%%%%%%%%%%%%%%%%%%%%%%%%%%%%%%%%%

\subsection{Code Generation}
\seclabel{codegen}
After \sys finds a summary for each input code fragment, the last step is to 
convert each such summary into a Hadoop task. The class encapsulating 
the Hadoop task has an {\tt execute} method, which takes as parameters all input 
variables in the code fragment. This method invokes the Hadoop task and returns 
an associative array that maps each variable identifier to its final value as 
computed by the Hadoop task. The associative array is then used to update the 
output variables before the remaining program is executed. Translation of $f_m$ 
and $f_r$ to concrete Hadoop syntax is done using syntax-driven translation rules. 
Since the postcondition is already in the MapReduce 
form, the rules to translate them into the concrete syntax of Hadoop are straightforward and 
omitted here for brevity.

\figref{full_example_mr} shows the final output code for the 3D Histogram example. 
{\tt HistogramHadoop} is the class generated by \sys, and the 
{\tt execute} method invokes the Hadoop runtime with the generated map and reduce 
classes. The resulting values---{\tt hR}, {\tt hG}, and {\tt hB}---are compiled and
returned by {\tt execute} and assigned to the original program's corresponding 
output variables as shown in \figref{full_example_output}. The code that 
reconstructs the arrays from key-value pairs is not shown for brevity.

\section{Evaluation}
\seclabel{evaluation}
We now describe our prototype implementation of \sys and present the
results derived from applying \sys to varied benchmarks.

\subsection{Implementation}
\seclabel{implementation}
\sys's program analysis and code generation modules are implemented by extending the open 
source Java compiler Polyglot~\cite{polyglot}. For synthesis, \sys uses an 
off-the-shelf synthesizer called SKETCH~\cite{sketch}. SKETCH uses counter-example 
guided inductive synthesis as its core algorithm.
The program analyzer encodes the verification conditions and search space in 
the SKETCH language. We implemented the functions and data structures 
required to model the semantics of MapReduce programs in SKETCH as well. 
In addition, \sys automatically models in SKETCH all program-specific user-defined 
data types. SKETCH performs bounded model-checking to generate a summary, which 
we then use to generate the Hadoop Code. We have not yet implemented \sys's formal 
verification component in \sys and therefore rely solely on bounded model-checking 
to verify correctness.

\subsubsection{Platform for Evaluation}
\seclabel{platform}
We used our \sys prototype to translate sequential Java benchmarks into Hadoop tasks. 
We measured the performance of both the original and the generated implementations 
on a 10 node cluster of Amazon AWS m3.xlarge instances. Each m3.xlarge node was 
equipped with High Frequency Intel Xeon E5-2670 v2 (Ivy Bridge) 2.5 GHz processors, 
15 GB of memory, and 80 GB of SSD storage. The cluster ran Ubuntu 
Linux 14.04 LTS, Hadoop 2.7.2 and Spark 1.6.1. We used HDFS for input data 
storage in both sequential and MapReduce implementations.

\subsubsection{Benchmarks}
\seclabel{benchmarks}
We evaluated the performance of \sys on the following five benchmarks. These 
benchmarks were taken from the Phoenix suite of benchmarks~\cite{pheonix}
and represent traditional problems that can be parallelized by rewriting
using the MapReduce paradigm.

\begin{itemize}
	\item The \textbf{Summation} benchmark sums all integer values in a 
	list.
  
    \item The \textbf{Word Count} benchmark counts the frequency of each word 
    in a body of text by iterating through each word in the input file.
	
  	\item The \textbf{String Match} benchmark determines whether a set of two strings is
	contained in a body of text. It returns a Boolean value for each string as
	output. Like Word Count, this benchmark also iterates through each word 
	in the input file.
	
  	\item The \textbf{3D Histogram} benchmark generates a three-dimensional 
	histogram that tallies the frequency of each RGB color component in an image (\figref{full_example_input}). 
	The output is an array for each color component that holds the frequency of 
	each intensity value.

	\item The \textbf{Linear Regression} benchmark iterates over a collection of
	cartesian points $(x,y)$ and computes a number of coefficients for linear 
  	regression: namely, $x$, $y$, $x*x$, $x*y$, $y*y$.
\end{itemize}

All benchmarks read input data from a text file saved on HDFS. For the generated
Hadoop solutions, class {\tt org.apache.hadoop.mapred.FileInputFormat} is used to 
read and split data across multiple mappers.

\subsection{Compilation Performance}
\seclabel{compilerperformance}
This section reports the time that \sys takes to generate Hadoop implementations 
and discusses the quality of these implementations.

\subsubsection{Scalability}
\seclabel{scalability}
\tabref{compiletimes} shows the average time (over 5 runs) required to synthesize 
a summary for each of the five benchmarks. \sys synthesized Hadoop implementations 
for all benchmarks within an hour. Simpler benchmarks, such as \textbf{Summation} and 
\textbf{Word Count}, were converted in under a minute and required only one 
iteration of grammar generation. No benchmark required more than two iterations 
to successfully synthesize an implementation.

\begin{table}[ht]
\centering
\begin{tabular}{c c c c}
\hline\hline                        
Benchmark 				& Program Analysis	& Synthesis and BMC & \# of Grammar Iterations \\ [0.5ex]
\hline                    
Summation 				& 	$<1$s 			& 13s	 			& 	1 \\   
Word Count 				& 	$<1$s 			& 44s				&  	1 \\ 
String Match 			& 	$<1$s 			& 1406s				& 	2 \\ 
3D Histogram 			& 	$<1$s 			& 2355s 			& 	2 \\ 
Linear Regression 		& 	$<1$s 			& 1801s 			& 	2 \\ [1ex]
\hline
\end{tabular} 
\caption{Average time for \sys to synthesize each benchmark.}
\tablabel{compiletimes}
\end{table}

\subsubsection{Sources of Parallelism}
\seclabel{parallelism}
A MapReduce program has two primary sources of parallelism. First, processing
can be parallelized in the {\em map phase} by partitioning the input data and spawning 
multiple mappers to process each partition simultaneously. Second, 
the {\em reduce phase} can be executed in parallel by grouping data to separate keys 
and aggregating for each key simultaneously. Hadoop also supports the use of combiners.
Before the shuffle phase, {\em combiners}---if used---aggregate data locally on 
every node to offer additional parallelism and decrease the amount of data that needs 
to be shuffled.

We now discuss the implementations \sys generated and how each leveraged both map 
and reduce side parallelism.

The \textbf{Summation} benchmark produces as output a single integer variable.
All data must be aggregated together and cannot be split to multiple
keys. The translated solution emits a key-value pair $(0, number)$ for each 
number in the input dataset during the map phase. These key-value pairs are 
aggregated locally on each node in parallel before being sent to the reducer. 
Note that key $0$ is the unique ID for the output variable.

The \sys-generated 
implementation of the \textbf{Word Count} benchmark emits $(word, 1)$ for 
each word encountered. The reducer then sums the values for each key. All 
nodes aggregate data locally (using a combiner) to compute word counts for the assigned data 
partition before the reducer aggregates intermediate results. In addition, \sys 
uses the words as keys. Therefore, the aggregation for different words is 
performed in parallel.

The generated \textbf{String Match} benchmark implementation parallelizes the search 
process. Each mapper iterates its assigned partition of text and emits 
$(key, true)$ whenever a key being searched is encountered. The data is locally 
aggregated by doing a disjunction of all values for a given key. Reduce side 
parallelism is leveraged as each key is aggregated in parallel.

The \textbf{3D Histogram} benchmark resembles the word count problem. Hence, the
\sys generated implementation iterates over
each pixel and emits $((color,intensity),1)$, where the key is a tuple 
of color and the intensity value. Data is aggregated in parallel in the reduce phase for each 
index of each histogram, for a total of 255$\times$3 keys. As with the preceding benchmarks,
data is locally aggregated before shuffling.

\textbf{Linear Regression} resembles the summation benchmark. All coefficients for
a given point ($x$, $y$, $x*x$, $y*y$, and $x*y$) are calculated and emitted by the mapper,
with a different key corresponding to each coefficient. 
For each key, the values are aggregated (by summation) locally before being globally reduced.

As is evident from all these benchmarks, \sys generated non-trivial 
implementations. \sys leveraged reduce side parallelism, reducing 
each output variable in parallel by assigning to each variable a unique ID and 
reducing data for each variable ID in parallel. For arrays, even greater parallelism 
was achieved by reducing each index of the array in parallel. \sys also exploited 
map side parallelism by evaluating expressions before they are emitted by the mapper
(e.g., as in Linear Regression). Lastly, \sys used the reduce class as a combiner 
to locally aggregate data whenever the reduce input and output key-value pairs 
were of the same type.

To evaluate the quality of optimization \sys achieved, we compared the runtime
performance of the original sequential implementations to the Hadoop implementations 
generated. We also examined the performance when synthesized summaries were 
manually translated to the Spark framework. Finally, to add context, we compared
the performance of Spark implementations generated by MOLD. \figref{charts} 
graphs the results of all five benchmarks against different dataset sizes.

\subsubsection{Alternate Implementations}
As discussed, \sys generates non-trivial implementations 
that effectively leverage the parallelism offered by Hadoop MapReduce. However, 
these implementations may not be the most efficient ones. 
%We now use the 3D Histogram 
%benchmark as an example to show how an alternate implementation may exist within 
%the defined search space but never found, as the search process in \sys stops as 
%soon as it finds a valid summary.
%
For the 3D Histogram benchmark, an alternative Hadoop implementation would be to 
emit for each pixel in the input data key-value pairs of the form $(intensity,color)$. 
Hadoop would then group the data by the 256 intensity values. Aggregation would
involve simply counting the number of times each color (Red, Green, or Blue) 
appears for a given key. Whether \sys generates this implementation or the one
discussed earlier in the paper depends upon which implementation the synthesizer 
discovers first. An important opportunity for future work is to enable \sys to 
use heuristics to reason about the optimal implementation.

\subsection{Performance of the Generated Benchmarks}
\seclabel{execperformance}

\begin{figure}
\centering
\begin{subfigure}[t]{0.325\textwidth}
\includegraphics[width=5.3cm]{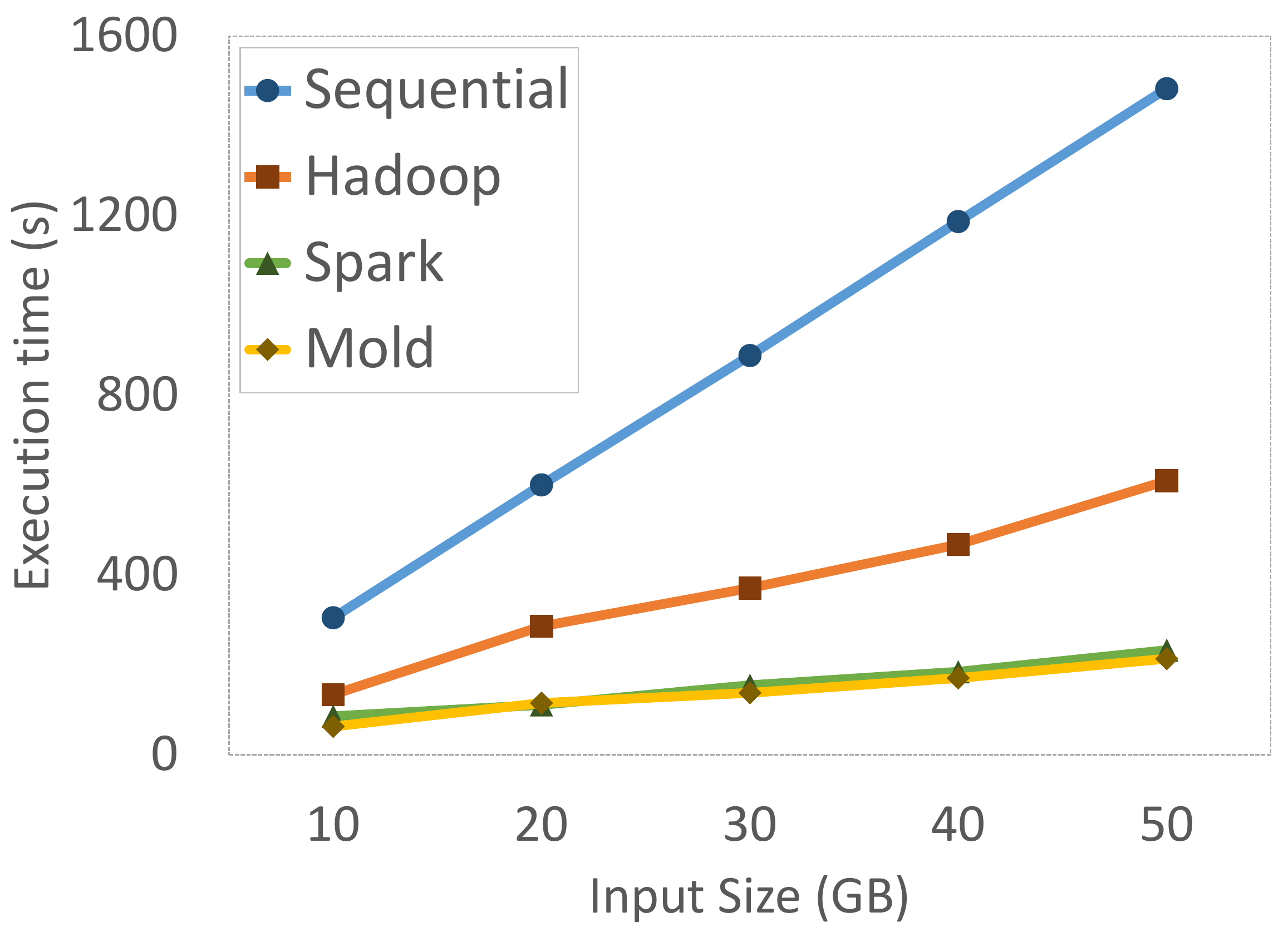}
\caption{Summation}
\figlabel{summation}
\end{subfigure}
\begin{subfigure}[t]{0.325\textwidth}
\includegraphics[width=5.3cm]{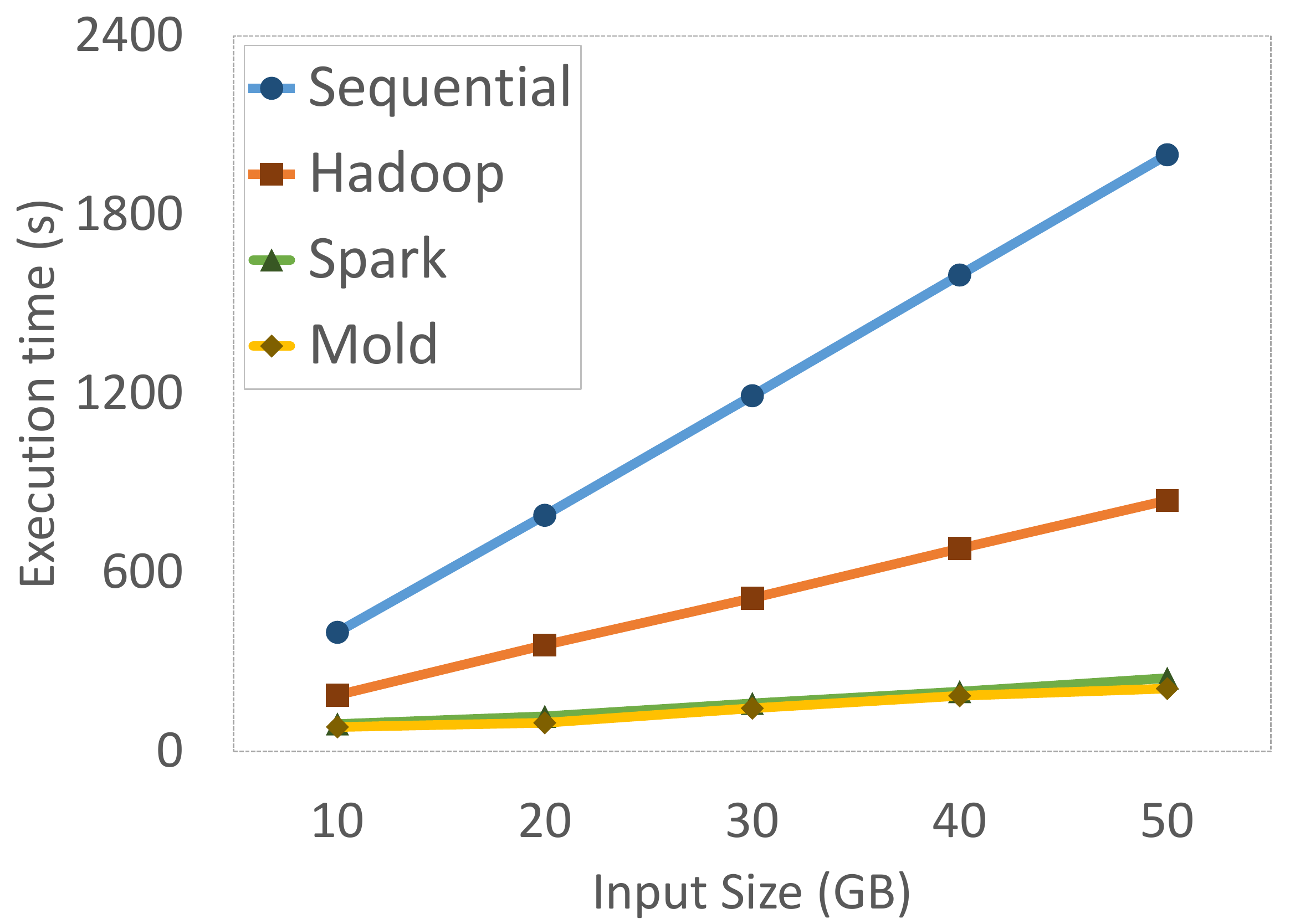}  
\caption{Word Count}
\figlabel{wordcount}
\end{subfigure}
\begin{subfigure}[t]{0.325\textwidth}
\includegraphics[width=5.3cm]{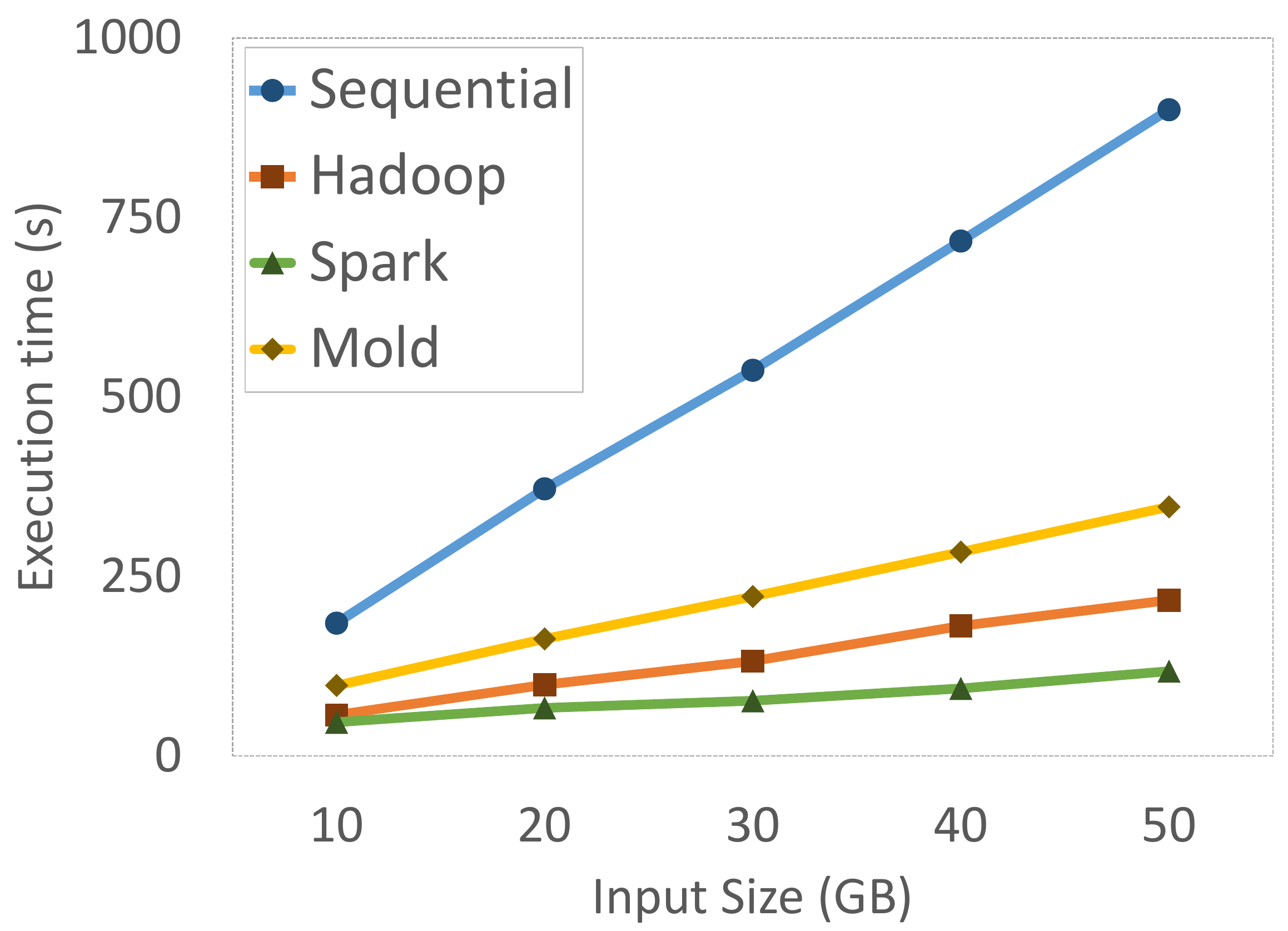}
\caption{String Match}
\figlabel{stringmatch}
\end{subfigure}
\\[0.2cm]
\begin{subfigure}[t]{0.325\textwidth}
\includegraphics[width=5.3cm]{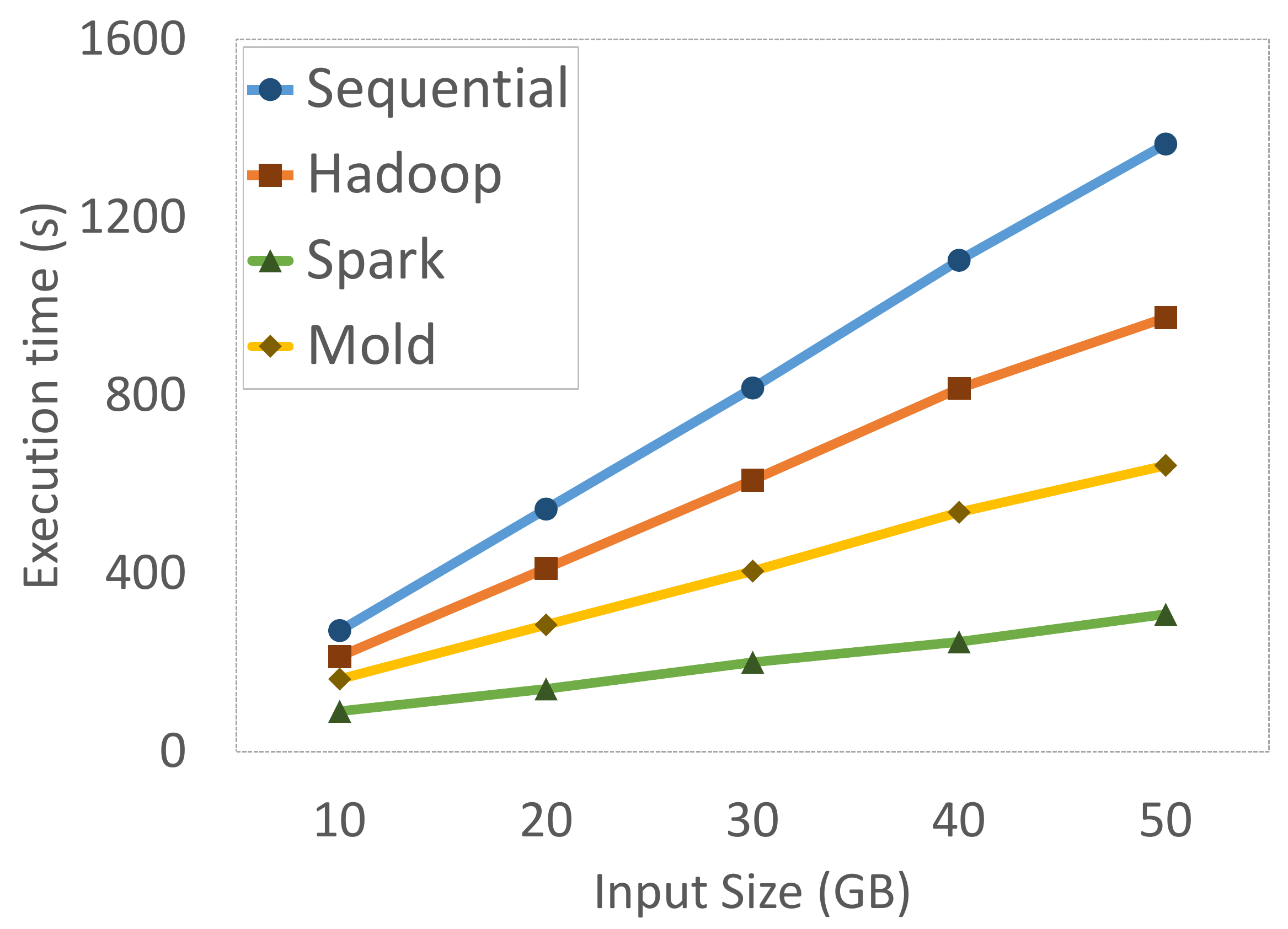}  
\caption{Histogram}
\figlabel{histogram}
\end{subfigure}
\begin{subfigure}[t]{0.325\textwidth}
\includegraphics[width=5.3cm]{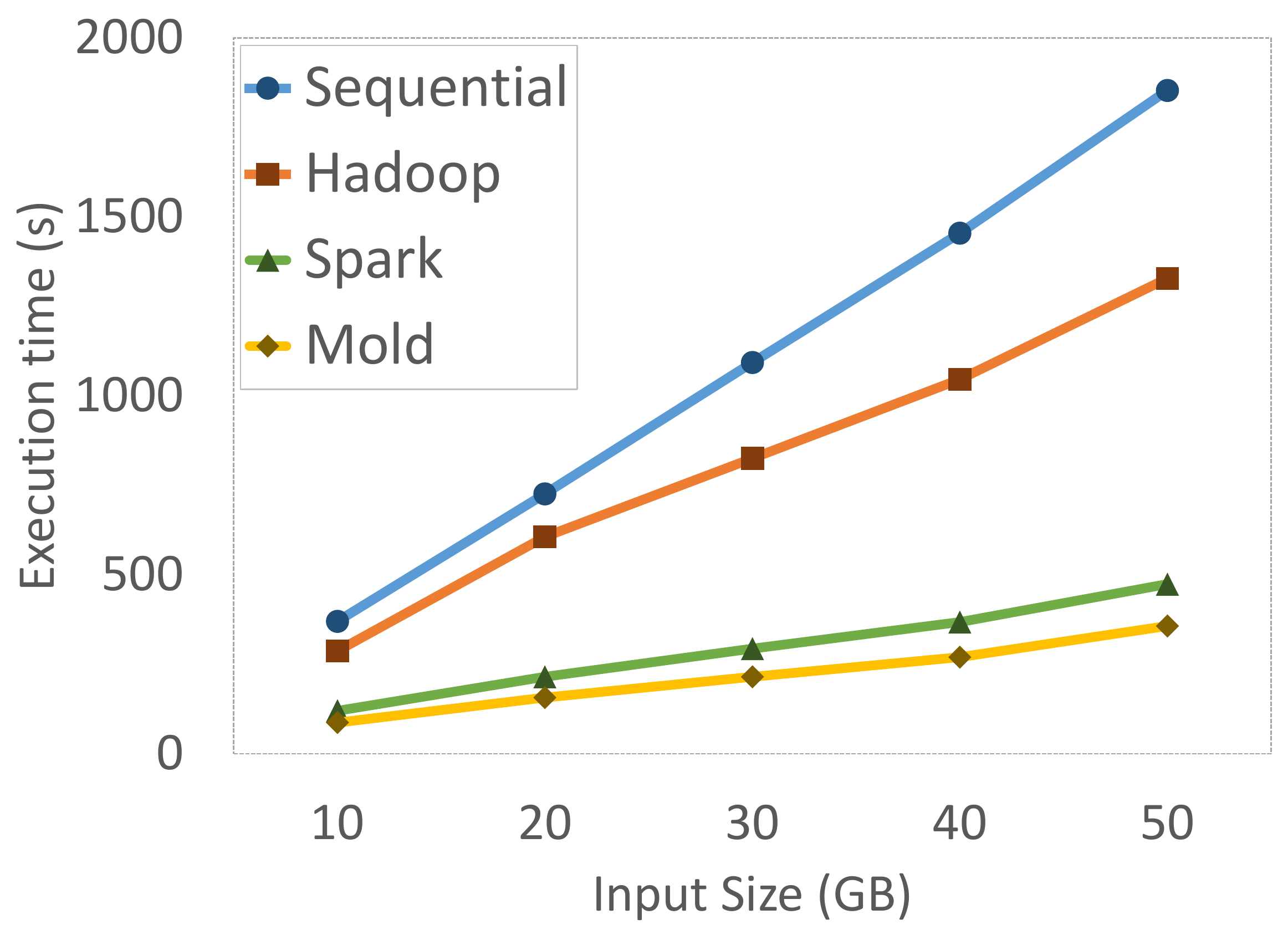}  
\caption{Linear Regression}
\figlabel{linearreg}
\end{subfigure}
\caption{Performance comparison of original implementations (blue) vs \sys optimized 
Hadoop (orange) and Spark (Green) implementations. Performance of implementations
when optimized using MOLD added for reference (yellow).}
\figlabel{charts}
\end{figure}

In all five benchmarks, the generated Hadoop implementations were not only faster 
than their sequential
counterparts but they also scaled better. Even for our smallest dataset (10GB), the 
Hadoop implementations outperformed the
original implementations. The average speed up for Hadoop implementations across 
all benchmarks was $3.3\times$ with a maximum speedup of $4.5\times$ in the case of 
String Match.

Translating the summaries synthesized by \sys into Spark yielded even higher 
speedups (up to $8.1\times$) since Spark uses cluster memory much more efficiently 
and minimizes disk I/O between different MapReduce stages. Extending \sys to 
automatically generate Spark code from the synthesized summary is currently a 
work in progress.

\section{Related Work}
\seclabel{relatedWork}
{\it \textbf{MapReduce DSLs.}} 
MapReduce is a popular programming model. It scales elastically, integrates well 
with distributed file systems, and abstracts away from the user low-level
synchronization details. As such, many systems have been built that compile code 
down to MapReduce~\cite{hive,spark,pig}. However, these systems provide
their own high-level DSLs in which the users must use to express their computation.
In contrast, \sys works with native Java programs and infers rewrites automatically.
\\
\\
{\it \textbf{Source-to-Source Compilers.}} 
Many efforts translate programs directly from low-level languages into high-level 
DSLs. MOLD~\cite{mold}, a source-to-source compiler, relies on syntax-directed 
rules to convert native Java programs to Apache Spark. Unlike MOLD, we translate 
on the basis of program semantics. This eliminates the need for rewrite rules, 
which are difficult to generate and brittle to code pattern changes. Many 
source-to-source compilers have been built in other domains for similar purposes. 
For instance, \cite{bones} evaluates numerous tools for C to CUDA transformations.
However, these compilers often require manual efforts to annotate the original 
source code. Our methodology works with code without any user annotation.
\\
\\
{\it \textbf{Synthesizing Efficient Implementations.}}
Extensive literature describes the use of synthesis to generate efficient implementations 
and optimizing programs. \cite{mrsynth} is the most recent research that attempts 
to synthesize MapReduce solutions with user-provided input and output examples. 
QBS~\cite{qbs} and STNG~\cite{stng} both use verified lifting and synthesis to 
convert low-level languages to specialized high-level DSLs for database applications
and stencil computations respectively.

\section{Conclusion}
\seclabel{conclusion}
This paper presented \sys, a compiler that automatically re-targets native
Java code to execute on Hadoop. \sys uses verified lifting to convert code
fragments in the original program to a high-level representation that can then
be translated to generate equivalent Hadoop tasks for distributed data
processing. We implemented a prototype of \sys and evaluated its
performance on several MapReduce benchmarks. Our experiments show that 
\sys can translate all input benchmarks, and the generated programs 
can run on average $3.3\times$ faster compared to their sequential counterparts.

\section{Acknowledgment}
The authors are grateful for the support of NSF grants CNS-1563788 and IIS-1546083 
as well as DARPA award FA8750-16-2-0032, and DOE award DE-SC0016260.

\nocite{*}
\bibliographystyle{eptcs}
\bibliography{references.bib}
\end{document}